\documentclass[acmsmall]{acmart}

\usepackage{booktabs}
\usepackage{multirow}
\usepackage{graphicx}
\usepackage{color}
\usepackage{tabularray}
\AtBeginDocument{%
  \providecommand\BibTeX{{%
    \normalfont B\kern-0.5em{\scshape i\kern-0.25em b}\kern-0.8em\TeX}}}

\setcopyright{acmlicensed}
\acmJournal{PACMHCI}
\acmYear{2024} 
\acmVolume{8} 
\acmNumber{CSCW1} 
\acmArticle{140} 
\acmMonth{4}
\acmDOI{10.1145/3637417}

\begin{document}

\title[Understanding Live Stream Meditation]
{Meditating in Live Stream:
An Autoethnographic and Interview Study to Investigate Motivations, Interactions and Challenges}

\author{Jingjin Li}
\authornote{Both authors contributed equally to this research.}
\email{jl3776@cornell.edu}
\author{Jiajing Guo}
\authornotemark[1]
\email{jg2263@cornell.edu}
\affiliation{%
  \institution{Cornell University}
  \city{Ithaca}
  \state{New York}
  \country{USA}
  \postcode{14850}
}

\author{Gilly Leshed}
\email{gl87@cornell.edu}
\affiliation{%
  \institution{Cornell University}
  \city{Ithaca}
  \state{New York}
  \country{USA}
  \postcode{14850}
}
\renewcommand{\shortauthors}{Jingjin Li, Jiajing Guo, and Gilly Leshed}

\begin{abstract}
  Mindfulness practice has many mental and physical well-being benefits. With the increased popularity of live stream technologies and the impact of COVID-19, many people have turned to live stream tools to participate in online meditation sessions. 
  To better understand the practices, challenges, and opportunities in live-stream meditation, we conducted a three-month autoethnographic study, during which two researchers participated in live-stream meditation sessions as the audience. Then we conducted a follow-up semi-structured interview study with 10 experienced live meditation teachers who use different live-stream tools. 
  We found that live meditation, although having a weaker social presence than in-person meditation, facilitates attendees in establishing a practice routine and connecting with other meditators. 
  Teachers use live streams to deliver the meditation practice to the world which also enhances their practice and brand building. We identified the challenges of using live-stream tools for meditation from the perspectives of both audiences and teachers, and provided design recommendations to better utilize live meditation as a resource for mental wellbeing. 
\end{abstract}

\ccsdesc[500]{Human-centered computing~Empirical studies in HCI}

\keywords{Mindfulness, meditation, live stream, well-being, autoethnography, first-person research}




\maketitle

\section{Introduction}

Mindfulness, a practice of maintaining awareness by bringing attention to the present without judgment \cite{kabat2009full}, has been growing in popularity in the past decade. Scientific research has shown that mindfulness practice brings various mental and physical well-being benefits such as reduced stress and anxiety, improved emotion regulation, improved focus, and reduced chronic pain \cite{brown2003benefits}. Similar to other cognitive or physical exercises, obtaining these benefits relies on consistent practice.  

\begin{figure}[h]
    \includegraphics[width=1\textwidth]{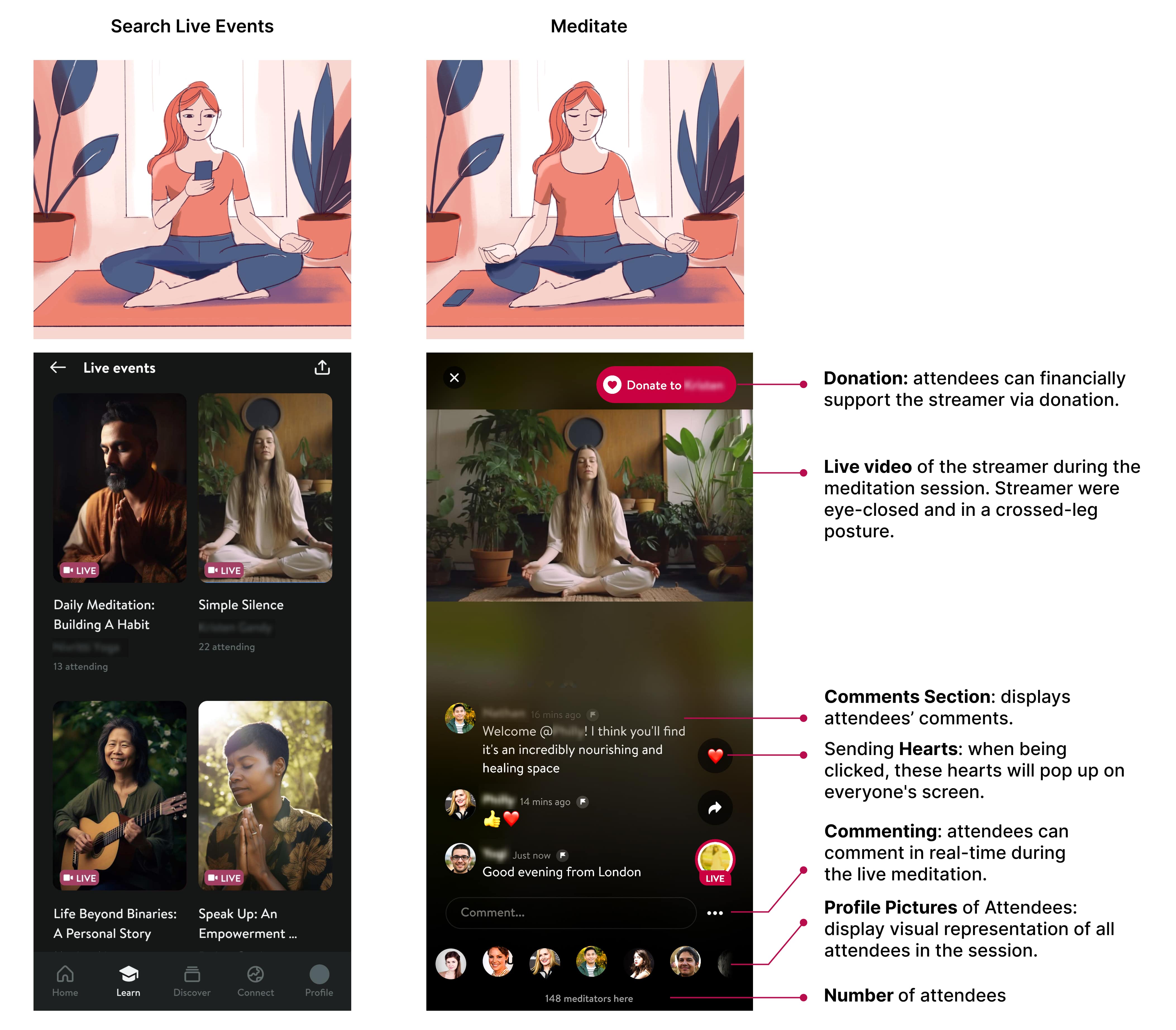}
    \caption{Two screenshots of Insight Timer live meditation. The left screenshot shows a list of available live meditation sessions. The right screenshot shows the interface during a live stream, with the components of donation, real-time video of the streamer, commenting, heart-sending, audience profile pictures, and a live count of attendees. The attendees' profile pictures have been replaced with copyright-free pictures for the purpose of anonymization.}
    \label{fig:insight_timer}
\end{figure}

Many technologies have been developed to support mindfulness, with a primary focus on teaching individuals how to practice meditation. These include, for example, mobile mindfulness apps (e.g., Calm, Insight Timer, Headspace), web-based mindfulness practice guidance \cite{krusche2013mindfulness}, wearables that detect stressful moments and deliver in-time mindfulness interventions \cite{kressbach2018breath}, and immersive virtual and augmented reality environments \cite{Tan2023-mindfulglasses, amores2016psychicvr, roo2017inner, prpa2018attending}. 
In addition to supporting individual mindfulness practices, technologies also harness the social aspect of mindfulness practice, such as group meditation via video chat and online communities for sharing information about mindfulness \cite{derthick2014understanding}, and a sense of social presence when practicing mindfulness in a VR environment \cite{Dollinger2021-hs}. Meditating in a video-based group has been found to be more effective in bringing state mindfulness and feelings of connection compared to meditating alone \cite{Hanley2022-rm}. Our paper further explores the role that social technologies play in supporting or facilitating mindfulness practice.  

In this paper, we look specifically at live stream technologies for practicing meditation with others, a practice that has been growing in popularity since the COVID-19 pandemic \cite{Da_Fonseca-Wollheim2020-nyt-meditation}. 
An example of live meditation interface is presented in Figure \ref{fig:insight_timer}. Live stream technology has been extensively explored in a variety of domains, such as gaming \cite{Hamilton2014-twitch}, performance \cite{Li2019-co-performance, Wang2019-greedy, Anjani2020-mukbang}, education \cite{Chen2021-live-education},
and informal knowledge-sharing \cite{Fraser2019-creative, Lu2019-cultural, Guo2022-fitness, Chen2020-towards, Faas2018-watch},
where the engagement between steamers and the audience plays an important role. However, we have a limited understanding of live stream mindfulness, where the attention of the audience is largely focused inward (e.g., breathing, body sensations) rather than on the streamer or the digital interface. 
The use of live stream tools in meditation practice might differ significantly from other live stream genres in terms of audience and streamers' motivation, interactions during the live session, and challenges that they encounter. We therefore explore the following research questions: 
\begin{itemize}
    \item RQ1: What motivates people to teach and attend live meditation?
    
   \item RQ2: How do people interact with each other in the live meditation sessions? 
   
    \item RQ3: How does live stream meditation differ from meditation in other formats (face-to-face, pre-recorded soundtracks) and other types of live streams?
    \item RQ4: 
   
    How do live streams support the practice of meditation and what challenges do people face?
\end{itemize}

To address these research questions, the first two co-authors conducted a three-month autoethnographic study to gain an in-depth, first-hand experience of attending live stream meditation as an audience. This was followed by a semi-structured interview study with 10 meditation teachers who use various live stream tools to understand their motivations, experiences, and perspectives on live-stream meditation. 

Our findings show that live stream meditation, although having a weaker social presence than in-person meditation, provides opportunities for attendees to build the practice routine and connect with other meditators. Teachers use live steam to deliver meditation to a global audience, which facilitates their own practice and brand building. 
We also identified specific challenges for both the audience and the teachers, of using live-stream tools for meditation practice. 
Audiences face the challenges of lacking clear information about the session, distractions from external and online environments, and finding a balance between practicing alone and attending live sessions. Teachers also experienced distractions while delivering mindfulness, along with the challenge of encountering anti-social behaviors, and issues related to receiving and requesting donations. 

Our primary contribution is an in-depth understanding of the motivations, interactions, and challenges of attending live meditation from both the teachers' and audience's perspectives, through a combination of an autoethnographic study and semi-structured interviews with meditation teachers. We also contribute to recommendations for redesigning and utilizing live stream tools to support meditation practices for mental well-being.

\section{Related work}

In this section, we first review previous work on mindfulness and various technologies that support it. Next, we review research on live streams for informal knowledge-sharing and companionship to give a broader view of live stream technologies and situate live meditation within this context. Finally, we review the first-person research on mindfulness HCI literature and justify our methodology choice of autoethnography and interviews. 

\subsection{Mindfulness Practice and Supporting Technologies}

Mindfulness has been commonly defined as paying attention to the present moment without judgment by Jon Kabat-Zinn \cite{kabat2009full}. It can be practiced through a variety of techniques, one of which is meditation. 
Although rooted in Buddhism, mindfulness has been adapted as a secular intervention in recent decades to help clinical and non-clinical therapy groups regulate their attention and emotions for mental and physical health-related outcomes \cite{Creswell2017-sk}, such as Mindfulness-based stress reduction (MBSR), mindfulness retreats and brief interventions, internet- and smartphone-based mindfulness interventions, etc.

The benefits of mindfulness, especially when practiced consistently, have been well documented in scientific findings such as reduced stress and anxiety, improved emotion regulation, improved focus, and reduced chronic pain \cite{brown2003benefits}. A growing number of HCI researchers and industry practitioners have been developing technologies to support people in learning and practicing mindfulness as a tool to obtain health outcomes \cite{terzimehic2019review}, such as mobile mindfulness apps (e.g., Calm, Headspace), web-based mindfulness classes \cite{krusche2013mindfulness}, wearable devices \cite{kressbach2018breath, Cochrane2022-zv}, and virtual reality environments \cite{amores2016psychicvr, roo2017inner, prpa2018attending}. One kind of mindfulness technology aims to detect the user's state and deliver timely mindfulness interventions through different modalities (e.g., audio, visual, haptics) \cite{Cochrane21, roo2017inner, Niksirat2019-yr, Sas2015-bn}. Another kind of mindfulness technology focuses on providing guided mindfulness sessions through websites \cite{krusche2013mindfulness}, commercial mobile applications (e.g., Calm, Headspace, Insight Timer), and immersive virtual reality environments \cite{Feinberg2022-lq, Dollinger2021-hs}. These technologies offer rich resources particularly for novices, guiding them in specific meditation practices\cite{lukoff2020ancient}. Previous research also explored the ways in which mindfulness practitioners incorporate various technologies to practice mindfulness and meditation into their daily routines, the benefits and challenges \cite{Markum20, derthick2014understanding, Li2022-beyond, Li2022-designing}, as well as the perspectives of mindfulness practitioners on possibility of future technologies in enhancing mindfulness \cite{Li2023-magicmachine, dauden2020body}.

While mindfulness can be practiced individually and is traditionally viewed as a self-centered practice, engaging in mindfulness with others offers unique advantages. For example, teachers offer guidance and exemplify a non-judgmental and compassionate stance \cite{Van_Aalderen2014-rw}. Peers contribute to the accountability of one's practice, the exchange of insights and perspectives and provide a sense of community \cite{Wahbeh2014-os, derthick2014understanding}. Prior work also suggests that video-based group meditation is more effective in bringing state mindfulness and feelings of connection compared to meditating alone, indoors or in nature \cite{Hanley2022-rm}. In response, several studies looked at technologies to support group meditation practice. 
For example, ZenVR \cite{Feinberg2022-lq} is a VR meditation learning system with a virtual pre-recorded teacher voicing the structured curriculum to a group of learners represented by avatars, mirroring an offline in-person meditation learning environment. 
Another example is Mindful Garden, an AR system that displays guided individual's biosignals as flowers to support co-located mindfulness meditation for greater connection \cite{Liu2022-wz}.

Video conferencing and live streams have become more common for practicing mindfulness after the outbreak of COVID-19 pandemic \cite{Li2022-beyond, Da_Fonseca-Wollheim2020-nyt-meditation, Chen2020-nyt-herculean}. 
In live meditation, streamers and other audiences are involved in one's practice, and are represented differently in various live-stream platforms (e.g., username, audio, videos). 
Despite its growing popularity, there is a lack of understanding regarding how individuals actively participate in live meditation and how these interactive elements influence the overall experience. Further exploration is needed to understand this phenomenon and the potential of utilizing live-stream tools in supporting mindfulness. 

\subsection{Live Streams for Informal Knowledge-Sharing and Companionship}
\label{bg:live-stream}

Live stream has become an increasingly popular interaction approach to connect people in the last decade. 
Live videos and real-time chat are typical modalities in live stream services \cite{Hamilton2014-twitch}. Streamers can share their screen or camera in the live video, and viewers can interact with the streamer and with peer viewers via text chat. 
Some platforms also support features such as virtual gifts \cite{Lu2018-china}, donations \cite{Wang2019-greedy}, or ad-hoc video chat with other streamers \cite{Li2019-co-performance}. 
A large body of work has studied live streams in the entertainment domain, such as gaming \cite{Hamilton2014-twitch}, performance \cite{Li2019-co-performance}, mukbang \cite{Anjani2020-mukbang}, and outdoor activities \cite{Lu2019-vicariously}. 
In recent years, HCI scholars started to explore live streams in niche domains, such as creative work \cite{Fraser2019-creative}, programming \cite{Chen2020-towards}, cultural heritage \cite{Lu2019-cultural}, companionship \cite{Taber2020-companion, Lee2021-study_with_me}, matchmaking \cite{He2023-seeking} and fitness \cite{Guo2022-fitness}. 

Informal knowledge-sharing is an increasingly popular genre of live streams such as cooking, skin-care, personal health, \cite{Lu2018-china}, creative art \cite{Fraser2019-creative}, and fitness \cite{Guo2022-fitness}. 
Without lectures or slides, streamers usually share knowledge or skills by demonstrating the procedure of an activity. Unlike mainstream live videos for which people watch to relax and kill time \cite{Lu2018-china}, knowledge-sharing streams usually have a smaller viewership (fewer than 100 average concurrent viewers) compared to gaming streams \cite{Fraser2019-creative}. Knowledge-sharing live streams are appealing as they \textit{"disseminate knowledge in a more relaxed, casual way than tutorials or lecture videos"} \cite{Fraser2019-creative}.
Viewers watch these kinds of streams with the purpose of learning and self-improvement \cite{Fraser2019-creative, Lu2018-china, Lu2019-cultural}, and the audience sometimes has different characteristics compared to gaming communities. Guided meditation can be seen as informal knowledge-sharing, as teachers share their meditation techniques by speaking out a script or demonstrating the practice to attendees who participate in the meditation. 

Challenges have been reported in informal knowledge-sharing live streams. For example, streamers found it hard to balance viewer engagement and the quality of their own activities \cite{Fraser2019-creative}, experiencing high cognitive demands when switching focus between the chat and their own work \cite{Fraser2019-creative, Lu2019-cultural, Guo2022-fitness, Alaboudi2019-live-programming}. 
In group fitness live streams, teachers are concerned about teaching quality and participants' safety, and participants find it hard to give real-time reactions while exercising at a distance from their laptops \cite{Guo2022-fitness}. 
Little research has been done about the practices and challenges of guiding and attending meditation sessions through live streams. 

Another special genre of live streams is companionship or co-working streams where viewers focus on their own experience instead of social interactions \cite{Taber2020-companion, Lee2021-study_with_me, Fraser2019-creative}. 
Viewers might not actively engage with the streamer or other viewers. Instead, they often focus on their own work or study and select various strategies such as covering the streamer's face or turning off sound to balance distractions from external ambience with the desire for companionship \cite{Lee2021-study_with_me}.
Given the uniqueness of meditation which emphasizes peace and inward discovery, it is unknown whether the motivations, interactions, experiences, and challenges are similar to those in other companionship live stream domains.

\subsection{First-person Research Methods and Research Positionality}

First-person research methods refer to \textit{"qualitative research approaches that turn to the researcher as the subject of inquiry."} \cite{Desjardins2021-co}. First-person research methods have grown in popularity in HCI and have been valued and adopted by many researchers to obtain a lived and embodied understanding of technologies or phenomena \cite{OKane2014-zx,Hook2010-nq,Homewood2023-fc,Lucero2018-mr,Claisse2023-uc, Gaver2023-yn, Nunez-Pacheco2023-nk, Wu-auto2024}. For example, Höök \cite{Hook2010-nq} used first-person method to study the embodied experience of horseback riding and translated the learned insights into design. Homewood \cite{Homewood2023-fc} employed an 18-month autoethnography of using self-tracking technology to mitigate long COVID contributing to the design of pacing technologies. 
Lucero \cite{Lucero2018-mr} presented an autoethnographic account of the experience of living without a mobile phone over 9 years. 

Autoethnography, a subset of first-person research methods, is an approach where researchers become participants in an ethnographic study to get a first-hand understanding of users’ everyday lived experiences \cite{Adams2021-th}. Autoethnography applies similar ethnographic techniques such as participant observation, but in contrast to ethnography that focuses on understanding the other (group or culture), autoethnography is intentionally self-referential, placing the researcher \textit{"as the primary character and author of a story"}  \cite{Poulos_undated-bf}. Autoethnography provides a unique perspective that embraces the subjectivity in the research:
\textit{"Autoethnography is one of the approaches that acknowledges and accommodates subjectivity, emotionality, and the researcher's influence on research, rather than hiding from these matters or assuming they don't exist."} \cite{Ellis2011-ii} Autoethnography also has the unique benefits of gaining empathy and understanding of users’ experiences by putting oneself in their shoes, and obtaining an understanding of nuanced experiences when studying users is difficult and out of reach (e.g., a hard of hearing individual \cite{Jain2019-uk}). 

Within mindfulness HCI research and related fields, first-person methods have been adopted to explore meditative experiences \cite{Cochrane21, Cochrane2022-zv, Mah2021-ia} and spiritual practice \cite{Claisse2023-uc}. For example, Mah \cite{Mah2021-ia} applied an autoethnographic approach to study the experience of cultivating compassion through the Buddhist meditation Tonglen. Cochrane et al. \cite{Cochrane21} developed an Inside-Out Probe workbook to study first-person walking meditation experiences with interactive soundscapes. Claisse et al. \cite{Claisse2023-uc} utilized autoethnography and interviews to study Buddhism practice during COVID-19 and design implications for online faith practices.

Live meditation involves both streamers and an audience. In our research, we chose to apply the autoethnographic method to understand the first-hand lived experience of attending live-stream meditation as the audience. Our autoethnographic study draws on the subjective and lived experiences of the first two co-authors who are both graduate student researchers and have prior experience with in-person meditation sessions but not with live-stream meditation. Both have a strong interest in mindfulness practice, with one (R1: Jiajing) having regularly practiced yoga for over seven years and the other (R2: Jingjin) having practiced mindfulness on and off for six years. Regarding live stream experiences, R1 has experience of attending fitness, music, and knowledge-sharing streams, 2-3 times a month, while R2 participates in live fitness, music, and shopping sessions, once every other week.

This decision was informed by the unique benefits of autoethnography, such as the integration of personal experience and reflexivity into the research process \cite{Rode2011-fq}, and practical considerations including privacy protection mechanisms on live platforms and the challenges of long-term traditional participant recruitment. Reflexivity enabled us to critically reflect on our own experiences, emotions, and analytical approach, enhancing accountability and transparency \cite{Rapp2021-rj}. 

Despite the unique advantages, autoethnography also presents challenges such as balancing the personal with the analytical \cite{Anderson2006-lu}, and researcher vulnerability, emotion, and self-indulgence \cite{Jones2016-pf, Lapadat2017-ethics}. Prior work has proposed ways to make a valid and successful autoethnography especially connecting personal narratives with wider social, political, and cultural meanings \cite{Jones2016-pf}, and practicing reflexivity to critically reflect on the researchers’ positionality and how their personal experiences shape the research \cite{Wall2006-ba}, detailed documentation and reflections \cite{Desjardins2021-co}. In light of these considerations, we carefully adopted the autoethnographic method in our research by 
(1) involving two researchers in the autoethnographic study to practice collaborative autoethnography \cite{Chang2016-zm},\textit{"a multivocal approach in which two or more researchers work together to share personal stories and interpret the pooled autoethnographic data"} to address the challenges from autoethnography as a solo practice \cite{Lapadat2017-ethics};
(2) having structured detailed documentation of our live meditation experiences and reflections, and practicing reflexivity in the research meetings and writing process; 
(3) combining it with an additional qualitative method: semi-structured interviews with meditation teachers to provide streamers' perspectives and experiences of live meditations, as being adopted in other HCI research \cite{Claisse2023-uc,Cecchinato2017-aa}. 

\section{Autoethnographic study: Method}

In order to gain an in-depth first-hand experience of live-stream meditation, the first two co-authors (R1, R2) conducted an autoethnographic study between June and September 2022. Both are graduate students with some meditation experience, as detailed in our research positionality in the related work section. During the period of the study, both R1 and R2 experienced manageable moderate work-related stress. Our personal experiences provided a critical background for the analysis and interpretation of the live meditation phenomenon. It is important to note that autoethnography carries inherent limitations related to the researchers' subjective experiences \cite{Jones2016-pf}. We emphasize that our intention is not to capture the breadth of diverse experiences that exist among different users, but to leverage the advantages of the unique personal insight and embodied understanding from autoethnography.

\subsection{Live stream platforms}
We participated in live-stream meditation sessions on three different platforms:

\begin{enumerate}
  \item Insight Timer is a free meditation app that offers meditation soundtracks uploaded by mindfulness teachers. A live event feature\footnote{\url{https://insighttimer.com/live}} launched around 2020 
  enables teachers to broadcast live meditation sessions through the app. Teachers can create live events on the app at least two days before the scheduled time. Users can browse available events in the app and join any sessions they are interested in. 
  \item Zoom is a popular video conferencing software. Some teachers offer small scale meditation sessions and share the Zoom link via email lists, social media, and personal websites.
  \item Discord is an instant messaging social platform that offers features such as voice calls, video calls, text messaging, etc. We joined one meditation server where online group meditations are often organized by server moderators and members. 
\end{enumerate}
Our choice of these three platforms was based on their popularity for meditation, video conferencing, and live streaming, respectively. Both Discord and Zoom allow the audience to turn on their microphone and camera. Insight Timer only allows the audience to send text messages, hearts, and donations. 

\subsection{Procedure}
In the first week, we explored the types of live meditations available including guided meditation, music, silent meditation, yoga, lectures, and Q\&A. 

We then attended sessions at least three times a week based on our personal interest and availability each week. In addition to live meditation, we also attended in-person sessions as we were interested in exploring the differences between live-stream and in-person meditation (RQ3). The length of the sessions we attended ranged from 20 minutes to 90 minutes. An example structure of live meditation included an introduction by the teacher, the meditation practice, and a following short Q\&A. Our eyes were generally closed during the meditation practice, as guided by the teacher: "Close your eyes if you feel comfortable or maintain a soft, unfocused gaze to the front". We had our eyes open at the beginning of live streams and after the meditation practice ended.

We took pictures or screenshots during streaming sessions, and wrote structured journal entries after each session, responding to prompts that were modified from the Inside-Out Probe workbook \cite{Cochrane2022-zv, Cochrane21} to document our lived experience data in detail. Our journal prompts were: (1) describe your thoughts and feelings pre, during, and post-session respectively. (2) describe where your attention was directed. (3) describe the flow experience, if any. (4) describe your experience of the sense of self in relation to the environment. (5) describe session difficulties, distractions, and refocusing strategies you used. (6) describe your feelings of the connections with the teacher and fellow meditators. We spent about 10-25 minutes writing post-session journals. 

\subsection{Data analysis}
\begin{figure}[h]
    \includegraphics[width=\hsize]{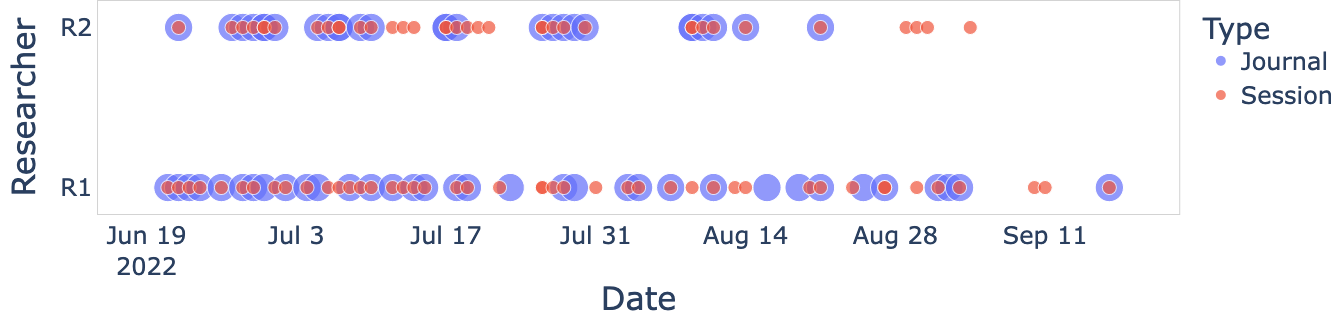}
    \caption{Visualization of attended meditation sessions. (Session: Researchers took brief notes and screenshots. Journal: Researchers wrote structured journals following the prompts in the workbook.)}
    \label{fig:journal_viz}
\end{figure}

We documented a total of 86 meditation sessions (R1:46, R2:40). Figure \ref{fig:journal_viz} presented the sessions that we attended and the journals we documented. In a few sessions, we took brief notes and screenshots instead of journals because these sessions were too short or the experience was too similar to previous sessions we attended. The journals and notes of these sessions were imported into Atlas.ti\footnote{\url{https://atlasti.com}}, a qualitative analysis tool after we removed personal details that we did not want to share. This allowed us to freely record our experiences initially, but also to preserve our privacy in cases of highly personal or sensitive records to have a balance between privacy and transparency \cite{Jain2019-uk}. 

We started analyzing the data as soon as we each had a few journal entries. Our analysis included open- and axial-coding \cite{saldana2021coding} with the following steps: 
\begin{enumerate}
  \item In the first round of analysis (open-coding), we each independently read the other researcher's journal and created open codes as brief comments that represented our interpretations of the other's journal entries. For example, we had comments such as "eyes open" and "focus on the voice guidance" that describe the physical and mental state during a session, "distractions from comments" and "distraction from external environment" to indicate experienced distractions, and "benefits of connecting", "challenge of finding a good session" that represent the benefits and challenges we had. 
  \item We then reviewed each others' codes, reflected on how our experiences were captured by the other researcher, and revised the codes that we felt were not interpreted accurately. 
  \item Next, we met to discuss and refine the codes together, and identified categories into which we grouped the codes (axial-coding). For instance, the comments "sharing personal stories" and "sending greeting messages" were grouped under the category "interactions during live meditation". 
  \item We performed the open- and axial-coding process iteratively several times, through weekly meetings, while attending live-stream meditation sessions and collecting more data. We stopped collecting data once we reached saturation and no new themes emerged. 
  \item Besides the analysis of the journals, during our weekly meetings, we also shared personal insights, experiences, and reflections from participating in live meditation and its impact on our mental well-being. 
\end{enumerate}

\section {Autoethnographic Study: Findings}
In this section, we report on our motivations to attend live meditations and the interaction patterns we observed during the session. We then compare the live-stream meditation session with other meditation mediums (soundtracks, in-person) and other types of lives streams based on our autoethnographic investigation. Lastly, we reported challenges and reflected on the personal benefits that we experienced in live-stream meditation sessions during the 3-month study.  

\subsection{Motivation to Attend Live Meditation}
At the outset of this study, we were driven to attend sessions as part of applying the autoethnographic method to obtain first-hand experience of live meditation. With time, we developed additional motivations to attend live meditation sessions: building a routine of self-care, taking a mindful break from work, connecting with other meditators, and the low cost of live sessions. These motivations are not mutually exclusive; often we had multiple motivations to attend a live-stream meditation session. Further, while some of these motivations are related to the unique aspects of the live-streaming environment (e.g., the low cost of attending), others are more general motivators to practice mindfulness (e.g., taking a mindful break). 

\subsubsection{Building a Routine of Self-care}
Building a routine of practicing mindfulness was our primary motivation to attend live meditation. While we are aware of the well-being benefits of practicing mindfulness consistently, forming this habit is not easy. During the three months-study, we started by trying a variety of live meditation sessions and settled down on several kinds of sessions that each of us attended regularly. R2 mostly attended morning sessions to start the day, while R1 preferred evening sessions after a day's work. 
Initially, R2 did not set reminders for the live meditations she wanted to attend and missed some sessions, and gradually started adding live events to her calendar. Scheduling live stream sessions made us more accountable for our mindfulness practice, as R1 wrote, \textit{"I feel more motivated to do meditations if I register for live events than opening a soundtrack. It’s like a promise to someone."} By registering for a live event, she felt obligated to attend, not to let herself (or the teacher) down by not showing up, thereby encouraging more consistent participation.

Interestingly, we didn't always strictly follow the teacher's instructions in live events. For example, R2 was writing a paper when a live session started and she didn't want to stop to attend the session. Instead, she played the live session in the background, treating it as an ambient mindfulness prompt, and she \textit{"tried to meditate for several minutes and went back to writing, trying to do it more mindfully."} In other words, sometimes the live stream session acted as reminders for us to tap into the present moment, turning our ongoing activities into informal mindfulness exercises. 

\subsubsection{Taking a Mindful Break from Work}
Besides building a routine, we found that attending live-stream meditation as a break from work was a recurring pattern in our experiences. We identified two types of breaks in our journals: spontaneous breaks and scheduled breaks. 

We both found ourselves spontaneously attending live sessions when we felt tired or mentally drained and in need of a break. R1 wrote: \textit{"After a long day's work I just want to take a break. I didn't book a session so I just browsed insight timer."} Attending live-stream meditation as a break helped her wind down her day and restore her energy. 

Spontaneous breaks served as an immediate remedy for our mental well-being, but we didn't always find the right session for us available at that particular time. For example, one time R2 needed a break to relax her body and mind, but only found movement-related sessions. She instead chose to follow a non-live guided body scan to meditate on her own.

As an alternative, R2 was more motivated to plan ahead and schedule live session breaks in advance, when she could browse through a variety of options to choose from. She wrote: \textit{"I scheduled this session 2h before it started because I wanted to use this session as a break from my work and it's definitely better than just aimlessly scrolling down social media."} 
We found that planning ahead helped us better manage our mental well-being, avoid overworking, and reduce the frustration of looking for appropriate sessions on the spot.

\subsubsection{Connecting with Other Meditators}
One important motivation for us to attend live stream sessions was to connect with other meditators, especially in sessions that we visited repeatedly. R1 experienced kindness from other meditators such as supporting comments and teacher's responses in live sessions, which motivated her to come back to these sessions, as she wrote: \textit{"I guess perhaps the nature of human interaction motivated me to keep meditating. Some people in the session, I never know who they are, but they are so warm and supportive."} This demonstrates that connecting with others, even strangers, provided R1 with a feeling of approval for her efforts to maintain the practice. 

Interestingly, although most live sessions we went to were led by a teacher, R1 also attended live self-organized group meditations on Insight timer and Discord. 
The organizers of these regular live sessions were members of a teacher's group. These sessions were structured as a practice with recorded guidance and informal chat among attendees rather than being delivered by a meditation teacher. Unlike teacher-led sessions, there was no introduction or Q\&A in these sessions. 
R1 found these self-organized small group sessions useful, even more than the teacher-led meditation sessions, because most of the session time was focused on the practice. She found it easy to build connections with other meditators because of the small group size and the recurring regular interactions. After attending a few Insight Timer live group sessions organized by other attendees, R1 wrote:
\begin{quote}
    \textit{I have joined this small group meditation 3 or 4 times. It's a really small group, only four or five people every time, but everyone here was so nice... I definitely am willing to keep coming to this group meditation.}
\end{quote}

As R1 pointed out, humans have a basic need for social connection and belonging \cite{maslow1943theory}. According to Operant Conditioning (OC) theory, an individual's behavior is shaped, maintained, or even discontinued based on the immediate and clear outcomes of that behavior \cite{skinner1965science}.
The effects of meditating can determine whether a person will continue practicing, and social connection can be seen as another layer of benefits that may reinforce meditation via live streams. 

\subsubsection{Low Cost of Attending Group Meditation}
R2 also attended in-person sessions besides live-stream sessions. Although she found that attending in-person group meditation practice with others increased her accountability for the practice, going to an in-person meditation was more costly in time and money. R1 pointed out that a live session just required her to attend with a phone or a laptop: \textit{"the cost is very low, I only need to put my phone aside. It is not a zoom meeting in which I should be in front of the camera. If I drop out in the middle, it is fine. It doesn't lose anything."} 
This echoes previous findings that the easy access to resources and the low cost of social presence motivate people to join live streams \cite{Lee2021-study_with_me, Guo2022-fitness, Lu2019-vicariously}.

\subsection{Interaction During Live Stream Meditation} \label{section:auto-social}

During sessions, we identified several patterns in our interactions with mindfulness teachers and other attendees.

\subsubsection{Non-verbal Cues}

We noticed that some of the physical spaces from where teachers were streaming included, for instance, calming wall art, lighting, home decorations, and inspiring quotes. Even though we had our eyes closed during most of the session, we appreciated the calming setup of these spaces. R2 frequently went to a session where the teacher used blue lights to set up her streaming background: \textit{"When I first entered the session, I was impressed by the instructor’s setup because the lighting was calming and pretty."} Such well laid out physical spaces provided a welcoming backdrop that was conducive for the meditation practice.

Audio quality is another non-verbal cue that can impact the quality of the live-stream session, but unexpectedly, we found that the perfect audio didn't always mean we had the best live session. 
Noise such as the sound of birds, traffic, wind, and even the teacher's breathing increased our perceived connection to the teacher, making us feel that we are in the same room. R1 wrote: \textit{"I can hear birds singing and the sound of traffic from the teacher's stream. I kind of view the connection between us even though we are across different time zones."} As we closed our eyes for meditation, the sounds coming from the teacher's side provided us with an ambience that created a physical presence illusion that strengthened our connection with the teacher. 

\subsubsection{Greetings and Goodbye Chats}
A common interaction pattern in live meditation is sending greeting and goodbye messages at the beginning and the end of a session. 
 
We observed that teachers responded back to these messages by addressing individual attendees.

Some meditators also greeted each other by calling out their names. 
 
As R1 wrote, \textit{"Many people said good morning and greeted each other. I think it's very sweet. I feel it's like you are in a retreat trip and in the morning people just get up and greet each other and start the meditation."} Greeting individual viewers are common practices in most live stream genres, especially small size streams, to build rapport and enhance the sense of connection \cite{Hilvert-Bruce2018-social, Guo2022-fitness, Hamilton2014-twitch}.

\textit{"I feel surprised to see people greet each other one by one. They are like old friends... Some people were even especially greeting me, even though I don't know them. This is so heartwarming."} R1 has attended other genres of live streams before, but she rarely saw strangers interact with such kindness. The community atmosphere motivated her to attend live meditation sessions as a routine. 
 
Sending out such messages also made R2 feel more involved in the live session than being a passive viewer, writing: \textit{"And I also said goodbye at the end to slightly participate and show my presence. I felt it was like a ritual compared to just listening to the soundtrack."} 
Interestingly, there was one Zoom live session in which the teacher asked everybody to unmute simultaneously to say hello at the beginning and say goodbye at the end, creating an opportunity for people to interact with each other at the same time. 

Taken together, these social experiences of greetings and goodbye interactions surround the individual inward focus of the meditation practice, providing a unique experience that cannot be achieved when practicing alone.

\subsubsection{Sharing Personal Stories}
In addition to greetings and goodbye messages, we also observed teachers and other meditators share their personal feelings and stories with strangers in the live session. Such sharing behavior often received supporting responses and encouragement from others.
R1 wrote: \textit{"After the meditation, the teacher said, you can comment here your difficult person... Someone said it’s her ex, someone named their parents. The teacher commented on the messages one by one. 
Her voice was so soft and the language was gentle. 
I appreciate that these people are willing to share their stories."}  
Self-disclosure has been reported in previous live streams studies. For example, a broadcaster shares their personal experience as a topic of live streams \cite{Lu2018-china}, friends met on Twitch develop friendship by intimate chat \cite{Sheng2020-irl}, or disclose personal information for specific needs \cite{He2023-seeking}. 
It is uncommon to see live stream audience self-disclose their personal experience and mental health issues openly in the chat, unlike in offline mindfulness groups \cite{Segal2018-qt} where confidentiality and safety are established via consent. 
Live streams can be viewed as an extension of existing interactions among mindfulness communities. 

\subsection{What's Different About Live Stream Meditation?}
\label{auto:diff}
Based on the analysis of our experiences, we identified several aspects in which live meditation is different from other forms of meditation, and from other live streams.
\subsubsection{Live vs. Soundtracks: More Accountability and Connection}
Before the beginning of this study, we both used recorded soundtracks to practice meditation if we needed guidance, found on various platforms such as meditation apps, YouTube, Spotify, etc. Unlike these resources, live meditation involves social interaction with the teacher and with other attendees (Section \ref{section:auto-social}). This social interaction motivated us to attend live stream sessions as \textit{"it’s like a promise to someone"} (R1), which increased our accountability. Further, interacting with others and especially when people were sharing personal stories, made us feel connected to them, something we haven't experienced when meditating with guided recordings.

\subsubsection{Live vs. In-person: Lower Cost and More Options}
Besides participating in live stream meditations, R2 also went to some in-person group meditations that required traveling and a monetary investment, which we found less convenient and and more costly compared to live meditations. In one case, R2 was debating \textit{"before the session, I was struggling with whether I should go to the session, it was a 10-minute walk from my workplace, and what if the session is not good? Should I stay there or I could leave."} The immediacy and accessibility of live-streamed meditation not only reduce the monetary and time cost, but also reduce R2's mental workload of weighing cost and investment. And the perceived ability to come and go as she wishes in the lives stream further lowers the social cost of staying and leaving.

Meanwhile, in-person sessions are generally limited by geographical locations and may not offer as many options as online sessions. Although R2 felt more involved in in-person sessions, she experienced greater freedom with online meditations, being able to switch to other sessions at any time if the session wasn't a good fit. However, this convenience came with a trade-off: R2 felt more natural to have informal conversations with other participants in in-person sessions, helping her feel more connected to them. Creating a similar sense of connection in live sessions required her to take the initiative to interact with others rather than to passively watch the live. 

\subsubsection{Live Meditation vs. Other Live Streams}

Before we started this three-month study, we had experience with other live streams for shopping, music, sports, and games. However, unlike other live streams, almost all interactions we experienced and observed in live-stream meditation were positive, with attendees showing kindness to each other. We believe that the audience of live meditation is unique, in that people come for the mental well-being benefits of meditation practice. 
Also, for most live meditation sessions, our eyes were closed and our attention was mostly toward our inner bodies (e.g., breath, body sensations), guided by the teacher's instructions. In contrast, in our previous live stream experiences, our eyes were open, focusing on screen-based activities and interacting with other people. Considering these differences, traditional live stream tools may not be able to support mindfulness sessions sufficiently.

\subsection{Challenges of Attending Live Meditation}

\subsubsection{Lack of Clear Information about the Session}
Unlike the recorded soundtracks with rich descriptions such as the type of meditation and scenario of use, most live meditations only provide a title and a short description that disappears after the session begins. They typically lack information about the meditation technique and session structure. 
 
As a result, we started some sessions not knowing what to expect, and it took us some time to figure out what was going on and if it was an appropriate session for us, as R2 wrote, \textit{"The meditation session was a bit confusing, at first, I thought it was body scan, but there was no full body scan"}. In recurring sessions, after we already attended and got familiar with the structure, we didn’t experience such challenges anymore.

Another frequently encountered information challenge was no clear indication of which point we were at during a session, especially when we entered a session that has already started, as R2 experienced, \textit{"I wasn’t sure what was going on with the session, and how long the session will be. I hope there were time stamps there so I would not waste so much time trying to figure out what point I was in the session."}
Sometimes we felt frustrated to join a live stream where we saw the teacher talking, we waited for several minutes and then the session ended. 
We then had to look for another live stream.

\subsubsection{Distractions}
Despite the flexibility and convenience of joining live meditation from anywhere, unlike controlled in-person group meditation, we experienced distractions from our physical and online environments. 

R1 often attended live meditation at home and her partner would chat with her while she was meditating. She had to stop the session to deal with the distraction, \textit{"I was interrupted by my partner. He asked me what I was doing. My mind shifted a bit but I quickly drew my attention back."} 
In contrast, R2 frequently joined the live session at work in a shared office. She used noise-canceling headphones to avoid distractions from people around her, which was effective in reducing the noise, as she wrote, 
\textit{"There were 3 people around (the office) while I was doing the meditation, [...] I was using my noise-canceling headphones, so I didn’t feel their presence and I was in my own world."} 
She was initially worried about others judging her meditating in the office with her eyes closed and legs crossed on the chair. This concern gradually disappeared as her colleagues got used to her habit and learned to avoid engaging in a conversation with her until after she completed the meditation and opened her eyes. 

We also experienced distractions from our online environments, such as phone notifications and other attendees' chat comments in the live session. Especially for R1, who occasionally opens her eyes due to boredom or to ensure her phone was still connected during silent meditation, seeing the chat comments interrupted her flow of meditation: \textit{"Unintentionally I saw some chats coming in. But the chat was too long. It was quite distracting to read that so I didn't read that chat and closed my eyes."} She put effort into moving her attention away from the chat and back to the meditation. 
In contrast, R2 was more easily distracted by the teacher, as she was used to practicing silent meditation. In one situation she had to lower the volume of the live session: 
\textit{"I was constantly distracted by the music because I would focus on the music instead of on my breath and my body sensations. So I lowered the volume."} 
This is similar to Lee et al.'s finding in "study with me" live streams \cite{Lee2021-study_with_me}, where the ambience can both motivate and distract viewers, and to fend off distractions people apply strategies such as flipping the phone, covering creators' faces, or turning off sound to create a study-friendly ambience. 

\subsubsection{Finding a Balance between Practicing Alone and Attending Live Meditation}
Though we found many benefits to attending live meditation, sometimes we found it more effective to practice on our own. For example, although there were various options for live meditation delivered by teachers worldwide, when we couldn't find the exact type of guidance we needed, or when we couldn’t resonate with the teacher’s guidance, we felt that it was better to practice alone or seek out other practice resources than attend live sessions.

For R2, especially in the first month of the study, she felt that live meditations helped her re-establish the habit of practice. However, she felt that it came with a reliance on these sessions to practice meditation, and she tended to only practice meditation when she found a suitable session. Recognizing that, she gradually started utilizing other resources, as she wrote, \textit{"(I was) Not able to find an ideal live session. So I went to search for several soundtracks, but I just felt like I didn’t want instructions, so I just used the timer. I set a timer for 20 minutes and I did breathing and body scan."} At this point in her mindfulness journey, R2 had the capacity to choose between live streams, recorded guidance, or self-guided meditation. Also, as presented in Figure \ref{fig:journal_viz}, we realized that the number of live sessions we attended decreased toward the end of the study. It was not only because we reached data saturation in our journals, but also because the live sessions scaffolded forming our practice habit and we were able to practice meditation more on our own. We needed less live meditation as we developed independence in our practice. 

\subsection{Reflections on Our Personal Benefits of Live Meditation}
\subsubsection{Strengthening Our Meditation Capabilities}
One common challenge we have in meditating alone is the difficulty of bringing attention back during mind wandering. Attending live meditation better facilitated redirecting our attention back. Both of us have experienced boredom, distractions, or the urge to do something else during meditation, but the voice and presence of the teachers in the live session helped bring us back to the meditation. For example, R1 noticed her impulse of doing something else during the live session, \textit{"but seeing her (the teacher) meditate sincerely made me feel guilty if I just stop and play the phone."}  The presence of the teacher reminded R1 of the reason she was doing this, and helped her bring her attention back to the meditation.

Besides the impact of the teacher, the presence of other meditators, and especially seeing others' faces, helped R2 to bring her attention back to the meditation session. She commented in her journal that on Insight Timer, without seeing other people's faces, she didn't feel the impact of their presence on her meditation practice. However, in a Zoom live session, there were other attendees with their videos on, so when R2 opened her eyes in the middle of meditation, she could see they were meditating with their eyes closed, which made her continue the meditation. The social presence of others, teachers and meditators, experienced through live meditation encouraged us to bring our attention back, toward building stronger meditation capabilities. 

\subsubsection{Building and Expanding Our Mindfulness Practice}
In addition to the immediate benefits of meditation such as becoming more peaceful and more clear-headed, attending live meditation served as a unique facilitator for building the habit of consistent mindfulness practice.
Unlike a digital reminder we set on our phone or laptop to practice meditation, with notifications that we found too easy to dismiss or postpone, committing to a live meditation session provided a time window that we had to actively either choose to attend or miss the session. R1, in particular, felt more invested in attending live stream sessions than meditating on her own because of the experiences of human connection, and this, in turn, helped her build a stronger meditation routine. 

Sometimes, as the time for a pre-registered session arrived, we weren't available or willing to start formal meditation delivered by the teacher, for example, when we were too engaged with work and found it challenging to pause for the session. These situations served as a prompt to incorporate mindfulness into our current activities. We found ourselves creatively using this time window to turn what we were doing into an informal mindfulness practice (e.g., R1: mindful walking, R2: mindful writing), with the live stream session in the background: 
\textit{"I just wanted to use this time window to do other activities mindfully. So I went back to my writing with the silent meditation in the background so I could hear the end bell" (R2).} 
Even when we didn't actively engage in formal meditation during every live session, the timing of these sessions nudged us to integrate mindfulness and apply awareness across a range of daily tasks, expanding our mindfulness practice beyond meditating. 

\section{Interview study: Method}
 
Our autoethnographic study provided us with deep personal insight into attending live meditation sessions. However, we lacked the perspectives of an important group of stakeholders: the meditation teachers. To bridge this gap, we conducted follow-up interviews with meditation teachers, which allowed us to answer our research questions from the perspectives of both the streamers and the audiences.

\subsection{Participants}
During the first-person study, we emailed the meditation teachers after attending their sessions, inviting them to participate in an interview. 
In total, we interviewed ten teachers (6F, 4M), with a range of teaching experience spanning 2 to 37 years. 
All of the participants offer live meditations for free, which means anyone can join their sessions without incurring a charge. They can receive donations from the audience if a donation feature is made available by the live stream platform. 
Eight participants work as full-time meditation teachers, offering paid one-on-one consulting, online courses, workshops, and retreats.
Two of them maintain a nine-to-five job and broadcast meditation on a voluntary basis.
Each interviewee received a \$20 donation in support of their work in the mindfulness community. The teachers follow a variety of meditation traditions, such as Pranayama,  Kundalini yoga, non-duality, and self-cleansing practice. 
Table 
\ref{tab:interviewees} shows detailed information of the interview participants. 

\begin{table}
\centering
\caption{Detailed Information of Interview Participants. YOE = Years of mindfulness teaching experience.}
\resizebox{\textwidth}{!}{
\begin{tabular}{lllllll}

\hline
\textbf{Pseudonym} 
& \textbf{YOE} 
& \textbf{Age Group}
& \textbf{Gender}
& \textbf{Platforms}
& \textbf{Country}
& \textbf{Job Type}
\\
\hline
Ryan & 37 & 65-74 & Male & Zoom, Insight Timer & US & Full time 
\\
Steven & 30 & 45-54 & Male & Insight Timer & UK & Out of interest
\\
Teresa & 10 & 45-54 & Female & Zoom, Insight timer, Facebook & US & Part time
\\
Chelsea & 2 & 55-64 & Female & ŌURA, Insight timer, Zoom & US & Full time
\\
Dave & 10 & 35-44 & Male & Twitch, Youtube, Zoom & US & Part time
\\
Ellie & 15 & 45-54 & Female & Zoom, Google meetings & US & Full time
\\
Brandy & 15+ & 45-54 & Female & Zoom, Youtube & US & Part time
\\
Sadie & 10 & 45-54 & Female & Zoom, Insight timer & IN & Full time
\\
Matis & 16 & 45-54 & Male & Insight Timer, YouTube, Twitch, Tiktok & FR & Full time
\\
Kelly & 7 & 25-34 & Female & Insight Timer, Zoom	& US & Part time
\\
\hline
\end{tabular}
}
\label{tab:interviewees}
\end{table}

\subsection{Interview Procedure and Data Analysis}
The interviews were conducted via Zoom from October to November 2022, each lasting between 50 and 58 minutes. The interview was structured into five sections:
 \begin{enumerate}
 \item First, after obtaining informed consent and providing an introduction, the interviewer posed questions about the teacher's background and their experience in practicing and teaching meditation.
 \item Next, we transitioned to discussing the teacher's experience and practice of guiding meditation via live streams: 
 when they started live streaming, what platforms they use, and how they schedule sessions. 
 We also asked about where they stream (e.g., outdoor, living room), the equipment they employ, and the audience of their live sessions. 
 \item Then, we delved into the interaction and connection with the audience before, during, and after the session, the comments they receive, the distractions that occur, and the interactions among attendees. 
 \item We then queried about the benefits and obstacles of guiding mindfulness via live stream, comparing it to in-person and recorded audio or video. 
 \item Lastly, we asked teachers to compare live meditation with other live streams (e.g., sports, games, music) if they had such experience.
\end{enumerate}

All interviews were audio-recorded with participants' permission and fully transcribed. Using an inductive open-coding analysis approach \cite{saldana2021coding}, the first two co-authors read through the transcripts multiple times and conducted the initial coding. 
For instance, we identified codes such as "benefits of expanding one's own practice" to describe a teacher's motivation for conducting live meditation, "energy in in-person sessions" and "convenience of live sessions" to highlight the differences between live and in-person sessions, "requesting donations" and "distractions from the message during the session" to describe the challenges that teachers reported. 
Then, both authors read through and discussed the coded transcripts together, iteratively highlighted excerpts, and identified key insights and recurring patterns in the data. We used pseudonyms (see Table \ref{tab:interviewees}) to label quotes from participants' transcripts.

\section{Interview study: Findings}

In this section, we report on teachers' motivations to offer live meditations, interactions, and the challenges they experience during live sessions. 
We also compare guiding meditation via live stream with other forms (e.g., in-person and recorded soundtracks). 
\subsection{Motivation to Offer Live Meditation}

In the mindfulness community, it is common to see teachers offer free classes as a social service in local communities \cite{Forte2019-yoga-senior-citizen}, typically in-person.
The outbreak of the COVID-19 pandemic prompted many offline gatherings to transition to online platforms \cite{nbc_2020-covid-ban-crowds, Chen2020-nyt-herculean, Ray2020-zoom-kitchen, Chen2021-live-education}, including mindfulness practice \cite{Da_Fonseca-Wollheim2020-nyt-meditation}, and the teachers who participated in our study expressed gratitude for the opportunity to continue teaching meditation, unbounded by geographic constraints.
Although in-person activities had resumed by the time of our interview study (October-November 2022) \cite{nyt-see-reopening, Wu2020-nbc-reopening-america}, 
all of our interviewees still broadcast live meditation at least once a week, stating they will continue offering meditation via live streams. The pandemic might have been a trigger for some teachers to start live streaming, but there are additional motivations to sustain this practice. 
In this section, we summarize the unique motivations of our interviewees for offering meditation via live streams, benefiting both themselves and others. 

\subsubsection{Enhanced Visibility and Availability of Mindfulness} 
Teachers use live stream tools to reach and deliver mindfulness to a wider audience beyond their immediate offline community. 
In his early career, Matis interacted frequently with children. 
He observed that the traditional meditation methods, which he had learned, were less effective for the young generation—described by him as \textit{'intoxicated with information'}—particularly those growing up with technology. 

\textit{"You need to find a way to seduce kids to catch their attention. It (live stream background) needs to be beautiful, it needs to be interesting."}
Following his exploration and research into streaming technologies, Matis evolved into a tech-savvy mindfulness teacher and streamer, aiming to make meditation more accessible and appealing to younger generations. 
He said,
\begin{quote}
    \textit{I come to offer people a different approach to meditation, or let's say, more people come to be interested in how to relax their mind... It's like an invitation to relax. And slowly people start to understand that meditation is necessary for themselves to take time, to quiet the mind, to have a look at themselves.}
\end{quote}

Matis's account demonstrates an interest in reaching out to people who are not in his immediate community to practice meditation in different forms, especially younger people who spend a significant amount of time online.
Another example is Dave, who in addition to leading mindfulness programs in corporations, 

produces meditation soundtracks on video and music streaming platforms, and offers live meditation sessions on Twitch and YouTube occasionally: 
\textit{"It's not about making money, but it is about hopefully putting content out there that will reach as many people as they can."} Dave is dedicated to educating people about the approaches and benefits of mindfulness practice, making the mindfulness practice more approachable, and spreading mindfulness to as many people as possible.

\subsubsection{Self-practice}
Living a mindful life is the pursuit of many mindfulness practitioners, including these live meditation teachers. 
Such a way of living requires awareness cultivated through daily informal and formal mindfulness practice. In addition to delivering mindfulness practice to other people, half of the teachers mentioned that sometimes they simply broadcast their daily practice. For example, 
\textit{"Basically that part of that is my everyday practice for myself. So I just opened it up and decided to do it every day for whoever wanted to join me"} (Ryan).

A fixed schedule of live sessions helped teachers keep their own meditation practice. 
Steven told us with a slight chuckle, \textit{"My main incentive to teach meditation is that it was the only way I'll be sure that I go to a meditation class and find the one who's actually teaching it..."} 
He indeed benefits from doing the live sessions regularly, as he explained,
\textit{"I think every time by the end of the session, I feel better and it's been worth doing and it's kind of forced me to do it in the time that I wouldn't normally."} The responsibility of teaching mindfulness via live streams provides teachers the opportunity to maintain their own mindfulness practice, which echoes our motivation of building a routine of practice by committing to others in our autoethnographic study.

\subsubsection{Brand Building and Marketing}
Full-time meditation teachers usually provide other paid services such as one-on-one consultations, online courses, workshops, and retreats. 
Live streaming on a regular basis helps them amplify their impact and establish their brand. 
For example, Sadie, a full-time meditation teacher located in India, utilizes live stream to expand the follower base: 
\textit{"When I did the live sessions, it was received very well. Suddenly the follower count increased. Listening was also increased. It was a big change"}. Live sessions also help teachers reach out to potential clients. Some attendees in Kelly's live sessions found her on other platforms and acquired  her services: 
\textit{"There have been people who have found me (on Insight Timer) and then come and follow me on Instagram, and then join my newsletter, and end up getting converted into a client or someone who comes on my retreat"}. 
While teaching live meditation is voluntary and free to the public for almost all our teachers, it provides them with an opportunity to build their personal brand and expand their business.   

\subsection{Interactions: Teaching Meditation via Live Streams}
\subsubsection{A Space for Streaming Meditation}
Most teachers are intentional about creating a physical space that conveys comfort, peace, and positive energy to the audience, mirroring our observations in the autoethnographic study. 
For example, Teresa has set up a dedicated meditation corner, with a yoga mat, LED lights, drums, and singing bowls. 
In addition to speaking her meditation scripts, sometimes she plays these instruments. She said, 
\textit{"I think when we do it in our special place as teachers, we really tap into the energy aspects. We tap into this neutral mind because you know, it's all through repetition... I want to build stability. They see my background. They know it's me."} 
Teresa meditates every day and she understands the consistency of persistence. She wants to spread the energy to her audience through the atmosphere and consistent practice.

Alternatively, some teachers prefer to make a space that is genuine and mirrors everyday environments. They recognize that noise or distractions are inevitable and may be viewed as part of the mindfulness practice. 
 
For instance, Dave wants to \textit{"embrace the noises"} as part of his meditation practice, and he wants his audience to hear what he hears, including ambient noise: 
\textit{"There's noise everywhere. When you're meditating, you're gonna hear noise... Some people might stop meditating and walk away, yeah, but we have to move through it, that's the whole point of meditating, right?"} Rather than eliminating outside distractions in his stream, Dave advocates for navigating through these distractions.

Moreover, most interviewees mentioned the concept of "energy" during meditation, particularly as it relates to the location from where they broadcast. 
For example, Kelly, a frequent traveler, will sometimes spontaneously go live on Instagram from beautiful places she visits. She says, \textit{"When I find a beautiful place, I just set my phone up, and go live with the title of `invitation to sit and meditate with me'. I just go and take some slow, deep breaths, and I don't guide or say anything. I just go sit and meditate and invite people into my space, basically, into that connection."} The beautiful setting with a tranquil though abstract energy, shared virtually with the audience, contributes to the aesthetic appeal of the sessions and fosters a deeper connection between the teacher and the audience.

\subsubsection{Chats and Hearts Features in Live Streams}

Similar to our autoethnographic experiences, the teachers reported that they start their live meditation sessions with a brief introduction to their theme or topic. They then lead 15-60 minutes of meditation practice, and conclude with answering questions or chatting with the audience. 
In addition to these verbal interactions with the audience, the live-streaming interface offers other ways to interact. All of our interviewees said they appreciate the "send heart" feature which gives them a \textit{"positive reinforcement"} (Dave). 
At the same time, Teresa voiced her concerns about the distractions created by the popping-up hearts on the screen, \textit{"while It's beautiful, but I think it's disrupting for many people. I think when you're mindful, nothing should bother you"}.

Similarly, the chat feature provides a channel for the audience to give feedback but it can also be distracting:  
\textit{"My eyes are getting pulled to the chat constantly, really huge kind of distraction"} (Ellie). 
Some teachers establish rules or make suggestions at the beginning of the session prior to the meditation part,  urging the audience o refrain from sending hearts or chatting during the meditation to minimize distractions. 
Other teachers close their eyes during the meditation practice to avoid distractions from the chat and other elements in the live stream interface. 
Some even mentioned opening their eyes from time to time to monitor the chat for potential technical issues.

Teachers who have utilized general live stream tools such as Twitch and YouTube reported that they did not use many features that were intended to increase engagement, such as emoticons and virtual gifts. As Dave explained, 
\textit{"They're great when you're live streaming (e.g., sports, gaming), you're having an interaction (with the audience), right? ... When you're meditating. Everyone's quiet, a lot of those features are going to go unused."} 
While these features are helpful in interaction- and engagement-focused live streams, they are not perceived as applicable or beneficial in the context of meditation sessions.

\subsubsection{Peer Interactions}
As we observed in the autoethnographic study, meditators have lots of interactions with others during live sessions. In our interviews with teachers, they also reported that their audience interacted with each other through chats. For example, Ryan observed his students talk to each other through comments, as he explained, \textit{"Based on what I see in the comments section, I think they do. They go, 'hey, hi, what's happening? What about this? What about that?'"} 

Most teachers hold positive opinions of such interactions among the audience. Brandy shared how people in her live Zoom sessions help each other: 
\textit{"this old lady who showed up yesterday, for the first time, and she was sharing her personal issues. And then the other woman who has some issues of her own was just corresponding and they were supporting each other. It was great."} 
Similarly, Matis considers chatting as a form of social support. 
\textit{"I know that at some points, someone would need to say something. If someone lost a parent or they're depressed, they will open up and different people will come to bring support."} 
Both Brandy and Matis appreciated platform features that allowed people to interact and share their personal stories with one another, especially when traumatic moments were triggered during the live session. 
The immediate responses people received provided compassion and support, similar to our observations in the autoethnographic study.

In contrast,  Ellie, who hopes the students focus more on the meditation practice, disables the Zoom chat in her live practice session, so students can only send private messages to her, 
as she explained, \textit{"Doing all this other stuff (chatting included) in the middle of a mindfulness session, it really diminishes the degree of presence that the participants in the program have access to, because the attention is split into three or four different modes. And that tends to cause distraction."} Disabling the chats minimizes the multi-tasking of students, so they can focus on meditation instead of interacting with others.

\subsection{How Does Teaching Mediation via Live Streams Differ from Other Communication Mediums?}
\label{interview:diff}
In our autoethnographic study, we compared our experiences in live meditations with in-person sessions and recorded soundtracks from the perspective of the audience. Here we report on the mindfulness teachers' perspectives on the differences between live meditation and other formats. 
Before we dive into the comparison, we provide some context about setting up live meditations.  

Nine interviewees reported that they previously had experience in guiding in-person meditation. They said that transitioning from in-person teaching to live streaming required extra effort, but they gradually figured out their own setup and workflow. 
First, being open to the global population across time zones, they needed to decide on the timing of their stream. Some teachers experimented with various schedules and ended up choosing times that worked for most of their audiences (Saddie, Dave). 
Alternatively, Ryan chose a time that fit his own preference of practicing mindfulness in the morning. Second, they needed to follow certain requirements to create live events, with rules varying between the platforms. While Twitch and Instagram only ask for the title of the stream, Insight Timer has more strict criteria (schedule at least two days in advance, high-speed internet) to start a live stream. 

\subsubsection{Live vs. In-person: In-person is Preferrable, But Live Creates More Opportunities}
Teachers unanimously favor teaching in-person group meditation, as it creates "collective energy" (an intangible feeling or atmosphere created by the group of people meditating together \cite{Vieten2018-ma}) and deepens the connection with other meditators.
Teresa explained how physical presence is important for her teaching: \textit{"I like to look at people, I like to see their eyes, I like to see how are they doing. Sometimes I see the person is yawning or shy like hiding... I want to give the modification, I want to help them."} 

In contrast, in most live-stream sessions teachers do not have the chance to dive deep and have interactions with every attendee because they do not have enough bandwidth to give customized feedback to everyone. 
Further, some participants find the lack of physical presence of the audience in live broadcasting more mentally demanding: 
\textit{"When the session was over, I was feeling tired, because I was missing this physical human connection to be with people face-to-face"} (Matis). 
It is worth noting that Matis sometimes uses specialized streaming hardware to go live on different platforms simultaneously, which requires him to process a lot of information at the same time. 
Despite the clear preference for in-person sessions, our findings point to a number of unique opportunities that live streaming offers to teachers. 

\textit{Reaching More People}.
Live meditations enable teachers to reach a broader audience worldwide. Ryan noted the global reach of platforms like Zoom, \textit{"The benefit of doing Zoom is I have people all over the world that show up at the same time. I mean, all over the world. And to me, that's magical."}
Brandy also remarked, \textit{"Even though I personally like in-person, it's only a small group. It's very local.. Like the university program, it's limited to that population. Whereas zoom, it offers to all the people in the world."} 
The increase in the scale of people they can reach gives teachers a sense of accomplishment, which motivates them to continue their efforts in delivering live meditation.

\textit{Virtual Environment Assists Meditation}.
Interface features in live stream platforms enable teachers to use visual cues for setting up an appropriate environment for meditation, such as a virtual background:  
\textit{"Now you've got clouds (demoing to the interviewer on Zoom). So I can use different backgrounds. And this is an advantage to be online. You can create an atmosphere, which comes to help people to be quiet, even come to watch"} (Matis).
Matis is skillful in live stream technologies and has applied virtual background, automatic slide shows, and camera switching into his live sessions, all with the goal of further engaging his audience in meditation practice.

\textit{Lower Costs}.
Full-time teachers face practical challenges of paying for a meditation studio and putting in the time and effort for marketing and managing a personal brand. 
Sadie explained: \textit{"It seems to be easier to connect with more people online, and especially for apps like Insight Timer, where they are taking care of the marketing, and teachers can focus only on teaching and being there."} As such, relying on a live stream platform reduces the burden of marketing and allows teachers to focus more on delivering mindfulness.

\subsubsection{Live vs. Soundtrack: Content Creation for Quality, Live Sessions for Consistency}
\label{section:}

Some teachers viewed the various media for delivering mindfulness as complementary, each having its own time and place. This is similar to our own experiences, where we found that soundtracks were a great substitution when suitable live meditations were unavailable. 
Ellie explained that live meditation together with other resources scaffolds a learner's mindfulness journey toward gradually learning to practice on their own without assistance from outside resources: 
\begin{quote}
     \textit{You've got apps where there's no contact. You've got live online sessions where there is contact and engagement, right? And then you've got practicing at home alone with no guidance at the far end of that continuum. [...] There's usefulness in all of those modes. I think if the intention is to support a person in establishing an independent practice that can last a lifetime. Supporting training toward few things, knowing oneself and one's needs, but also getting comfortable with practicing without input. It's important.}
\end{quote}

From the teachers' side, there is a trade-off between recording guided soundtracks and going live. Recording guided meditation soundtracks allowed them to control the quality of their content through repeated recording and editing. 
Further, by creating and posting more soundtracks, the number of their followers gradually grew. 
However, the pressure to consistently create high-quality content might lead to burnout \cite{Lorenz2021-nyt-creator-burnout}. 
Dave, having released four meditation albums, said that he started to feel pressure from followers who ask for new soundtracks. He sees live meditation sessions as a good alternative: 
\textit{"It's more important to put content out there that speaks to people. And I'd rather do that once a month, instead of doing it every week and putting content out there that doesn't speak to people. What I find is better on a continuing basis are the live meditations."}. For Dave, offering live meditations reduces the pressure to produce high-quality recorded content, and the archived live sessions still get views even after the live sessions have ended.

\subsection{Challenges of Guiding Meditations via Live Streams}
\subsubsection{Distractions from Technologies}
Echoing what we experienced in attending live meditation as an audience, live stream features such as chats and hearts were found to be distracting for teachers. 

To avoid these distractions, teachers employ various strategies such as closing their eyes or treating them as a mindfulness challenge. 
Teresa, for instance, expresses the disruption from the screen, 
\textit{"I love the hearts. It's beautiful. But I think it's disrupting for many people. I think when you're mindful, nothing should bother you. Nothing should be of importance. Like the phone should be off. Nobody needs to look at their texts or emails... But are they really truly present or they're looking at something else?"} 
Teresa, like other teachers, 
is concerned not only with her own distractions during the session, but also with the audience's potential distraction due to their devices

To maintain a balance between audience engagement and a distraction-free meditation, we identified teachers establishing specific rules for their live sessions. 
For example, Ellie requests that participants in her Zoom meditation sessions mute their microphones, keep their video on, turn off self-view to avoid looking at themselves, and avoid using the chat feature during the practice. 
  
Similarly, Sadie asks her Insight Timer audience to hold off on clicking the `heart' button during meditation to maintain focus.
Interaction can resume post-meditation, a more appropriate time, as she explained, 

\textit{"Once we end the meditation, I ask people for feedback. So that time is a good time, you know, to use the hearts or to share their feedback. "} These social rules help fend off distractions from interactive features, while taking advantage of them at more appropriate times.

\subsubsection{Anti-social Behaviors}

Though anti-social behaviors are rare in meditation live streams, they do occur. 
Most of our interviewees reported that they have encountered anti-social behaviors at least once, such as Zoom bombers, trolls, and spammers. 
Brandy said,
\textit{"We don't get a lot of bombers. But when we do get them, they're very disruptive. I hope there's some kind of security feature. We do have a waiting room, but I can't tell who's who, you know, because the bombers will always change their name each time. That's a really tough part."} With no physical presence, and the ease of creating new accounts, teachers find it challenging to keep trolls away. 

In most cases, teachers choose to ignore troublemakers and "let it go". Kelly explained, 
\textit{"Occasionally though, you'll get a troll in there. [...] They're just trying to get attention, throwing profanities out there, or just saying weird things. So you just ignore them. It's a practice of mindfulness, right. It's like, what am I choosing to focus on right now."}. 
By not responding to these anti-social behaviors, teachers avoid fueling such negative energy.
Further, ignoring disruptive behaviors is seen as a form of practicing focused mindfulness at a higher level.

\subsubsection{Receiving and Requesting Donations}
While most live meditation sessions are free to attend, all of our interviewees appreciated receiving donations from the audience. 
The audience can directly donate to the teacher through the platform during the live session, and some platforms support donations even after the session ends. 
Half of our interviewees work as full-time live meditation teachers, and they report putting a lot of effort to prepare, \textit{"(You need to) plan your day, be ready, you'd wake up at a certain time..."} (Sadie).
Donations give teachers instrumental and emotional support to continue offering live meditations.

While platforms like Insight Timer encourage teachers to mention donations two or three times during the live stream\footnote{Insight Timer. How to Host A Great Live Event: 
\url{https://help.insighttimer.com/support/solutions/articles/67000664796-how-to-host-a-great-live-event}}, 
directly asking for tipping might hinder the audience from doing that \cite{Lee2019-gifts}. 
Further, three interviewees expressed discomfort with this practice, as they do not want to give the audience the impression that they value monetary rewards over the mindfulness practice itself. 
Chelsea shared her challenge with directly soliciting donations during sessions, 
\begin{quote}
    \textit{The donation is really challenging for me... It has always felt clunky to me, although I've gotten better at it. Yesterday I had somebody donate, and I was like, Oh, thank you so much for your donation. And then I could tell them [...] what those donations are for, why it's beneficial, instead of just throwing it out there.}
\end{quote}

Explaining the purpose of the donation helps the teachers to relieve the tension between the nonmaterial meaning of mindfulness practice and the realities of living a material life. Some teachers suggested that the platform should be more proactive on donations, instead of pushing the job onto the teachers. 
\textit{"So probably a tier model where you have this many sessions free, but anything more than that, if you pay something, I think it would help the teachers"} (Sadie). 
Teachers would like to continue offering meditation guidance through live streams, and relying on the generosity of their audience poses an ongoing challenge, that they think the platforms can mitigate.

\begin{table*}[htb]
\centering
\caption{Summary of findings from the autoethnographic study and the teacher interviews: 
motivations for participating in live meditation (RQ1), interactions (similar across both studies) (RQ2), and challenges with corresponding strategies presented where applicable (RQ4). 
See Table \ref{table_difference} for RQ3 summary.}

\resizebox{\linewidth}{!}{%
\begin{tblr}{
  cell{2}{1} = {r=4}{},
  cell{6}{1} = {r=2}{},
  cell{6}{2} = {c=2}{},
  cell{7}{2} = {c=2}{},
  cell{8}{1} = {r=3}{},
  hline{1-2,6,8,11} = {-}{},
  hline{3-5,7,9-10} = {2-3}{},
}
                       & \textbf{First-person Attendees}                                                                                                                                                                                                                                                                                                                                                                   & \textbf{Teachers}                                                                                                                                                                                 \\
{\textbf{Motivations}\\\textbf{(RQ1)}}    & Building a routine of self-care                                                                                                                                                                                                                                                                                                                                                      & Self-practice                                                                                                                                                                                     \\
                       & Taking a mindful break from work                                                                                                                                                                                                                                                                                                                                                     & Enhanced visibility and availability of mindfulness                                                                                                                                               \\
                       & Connecting with other meditators                                                                                                                                                                                                                                                                                                                                                     & Brand building and marketing                                                                                                                                                                      \\
                       & Low cost of attending group meditation                                                                                                                                                                                                                                                                                                                                               &                                                                                                                                                                                                   \\
{\textbf{Interactions}\\\textbf{(RQ2)}} & {Non-verbal cues\\\labelitemi\hspace{\dimexpr\labelsep+0.5\tabcolsep}Physical set-up, e.g., calming wall art, lighting, home decorations\\\labelitemi\hspace{\dimexpr\labelsep+0.5\tabcolsep}Audio quality\\~}                                                                                                                                                                       &                                                                                                                                                                                                   \\
                       & {Chats and sending hearts \\\labelitemi\hspace{\dimexpr\labelsep+0.5\tabcolsep}Greeting and goodbye messages before and after meditation\\\labelitemi\hspace{\dimexpr\labelsep+0.5\tabcolsep}Hearts poping on the screen\\\labelitemi\hspace{\dimexpr\labelsep+0.5\tabcolsep}Sharing of the personal story\\\labelitemi\hspace{\dimexpr\labelsep+0.5\tabcolsep}Peer interactions\\~} &                                                                                                                                                                                                   \\
{\textbf{Challenges}\\\textbf{(RQ4)}}   & {Distractions \\\labelitemi\hspace{\dimexpr\labelsep+0.5\tabcolsep}From the physical environment, phone notifications, chats\\\labelitemi\hspace{\dimexpr\labelsep+0.5\tabcolsep}Strategy: noise-canceling headphones, lower the volume}                                                                                                                                             & {\\\labelitemi\hspace{\dimexpr\labelsep+0.5\tabcolsep}From chats and hearts\\\labelitemi\hspace{\dimexpr\labelsep+0.5\tabcolsep}Strategy: close eyes, set rules, disable peer chats} \\
                       & {Lack of clear information about the session\\\labelitemi\hspace{\dimexpr\labelsep+0.5\tabcolsep}E.g., meditator technique, structure, indicators}                                                                                                                                                                                                                                   & {Anti-social behaviors\\\labelitemi\hspace{\dimexpr\labelsep+0.5\tabcolsep}Strategy: remove and not engage with disruptors}                                                                      \\
                       & {Balancing between practicing alone and attending live\\\labelitemi\hspace{\dimexpr\labelsep+0.5\tabcolsep}Alone: more flexible and resources;~\\\labelitemi\hspace{\dimexpr\labelsep+0.5\tabcolsep}Live: more social accountability}                                                                                                                                                & Receiving and requesting donations                                                                                                                                                                
\end{tblr}
}
\label{tab:SumFindings}
\end{table*}

 \section{Discussion and Design Implications}
Our autoethnographic study and interview study uncovered many novel aspects of meditation live streams. Our key findings are summarized in Table \ref{tab:SumFindings} (RQ1, RQ2, RQ4) and Table \ref{table_difference} (RQ3). 
In this section, we first discuss the roles that social interactions play in live meditation. We then discuss and propose utilizing live meditation to maintain mental well-being. Next, we compare our findings about live meditation to other live stream genres research. We conclude with design recommendations for live stream meditation. 

\subsection{The Social Dynamics of Live Meditation: Opportunities and Challenges}
Prior work has shown that meditation practice has positive impacts on mental health \cite{brown2003benefits, Grossman2004-vf}. However, establishing a consistent meditation practice can be challenging. 
Our personal experience of participating in live meditation through an autoethnographic study over three months helped us build a routine of self-care and expand our mindfulness practice. 
Specifically, the social aspects of live streams encouraged us to attend the sessions and build connections with others, making us accountable to the practice. This finding echoes previous research that emphasizes the importance of support from peers and teachers in mindfulness practice \cite{Van_Aalderen2014-rw}. 
Similarly, the teachers we interviewed also identified the wide reach of live streams as an opportunity to help others get started, and for themselves to maintain consistent practice through commitment. 

Our findings suggest that live-streaming effectively translates the collective energy that both the teachers and we experience during in-person meditation sessions to an online setting, creating a supportive digital "container" \cite{lukoff2020ancient} that could be especially beneficial for newcomers to meditation. 
For instance, greetings or check-in messages from the teachers and other attendees help foster a welcoming environment. 
Real-time guidance and a collective presence offer a sense of companionship as opposed to that of solitary meditation. 
These social elements may facilitate the learning phase for beginners and sustain the practice for experienced meditators, which aligns with previous research that experienced meditators utilized online video conferencing tools for group meditation sessions \cite{derthick2014understanding, Li2022-beyond}. 

At the same time, both the teachers and we identified distractions during live meditations as a major challenge. 
While social interaction is important for physical and mental health \cite{Umberson2010-social-health}, especially during meditating, one is expected to focus inward quietly, and engaging in social chatting can potentially undermine the effectiveness of the practice. The challenge of distractions during live-stream has also been reported in studies of other genres \cite{Guo2022-fitness, Lu2019-cultural}.

To mitigate distractions during live meditation, we and the teachers utilized strategies such as closing our eyes, lowering the volume, utilizing noise-canceling headphones, and limiting social chat to before and after the meditation. 
Besides these strategies, we propose the design of a non-distracting live mediation interface in our design implications (Section \ref{design:non-distracting}) to harness the social aspects of live meditation and minimize distractions. 
Furthermore, while closing our eyes is one way to minimize screen-based distractions, exploring other engagement methods in meditation such as haptic \cite{Costa2019-ar}, textile \cite{Mauriello2014-ep}, and temperature-based \cite{Dauden_Roquet2021-gk} or even aerial feedback could bring the presence (e.g., breathing) of others, while still assisting in focusing on the inner body during meditation.

In addition to managing distractions from chats, hearts, and other social interactions, the social dynamics of live-streaming introduced additional unique challenges for teachers. 
They occasionally needed to handle anti-social behaviors \cite{Seering2017-shaping}, adapt to the monetary incentive mechanism in live streams, and received less feedback compared to in-person sessions. 
Though donation and gifting are common in live streams \cite{Lee2019-gifts, Wang2019-greedy}, requesting and receiving donations is rarely identified as a challenge. 
The lack of feedback from attendees in live streams was reported to cause anxiety and cognitive burden for teachers \cite{Guo2022-fitness, Chen2021-afraid}.
 
Both teachers and live stream platforms can endeavor to handle the challenges from the social aspect of live streams. 
For example, teachers can consider recruiting moderators from regular attendees to help them manage the stream \cite{Wohn2019-volunteer, Cai2021-moderation}, and they can set up community rules to align their preferences with audiences' expectations \cite{Cai2022-coordination}. 
Live stream platforms can offer designs that adapt to meditation streams. We elaborate in Section \ref{discussion:redesign}.

\subsection{Live Meditation as a Resource for Mental Well-being} 
Reflecting on our findings, we discuss here the best practices for utilizing live meditation as a resource to maintain well-being practices. We believe that these insights can be useful for other types of therapeutic and mental well-being exercises beyond meditation.

\textit{Make time for well-being}.
Previous research has suggested that busy lifestyles and lack of routine negatively affect the frequency of mindfulness practice \cite{Laurie2016-ud}. 
We found that scheduling live meditation in our calendars encouraged us to set aside the time, go to the session, and practice with others. 
Besides, having the live meditation on our schedules also served as a time for a mental check-in if we were unable or unwilling to stop our current tasks to attend the live session.  
Most live stream platforms can automatically add live events to users' calendars, with an immediate link to access the live session at the scheduled time. Previous work has shown that calendars are useful not only for managing one's schedule and tasks, but also as a source of reflection on one's identity \cite{Leshed2011Lie}, intimate relationships \cite{Thayer2012Love}, and health \cite{Barbarin2015Taking}. 

\textit{Scaffold the practice without over-dependence}. 
While attending live meditations offered us unique, real benefits, over time we gradually developed a habit of using other meditation resources (e.g., recorded soundtracks, a digital timer) to practice meditation on our own. Some of the teachers also reported being motivated to introduce new meditators to the practice, and help them develop their self-practice beyond live meditation. 
We suggest that practitioners in mental well-being areas could utilize live streams as a digital container \cite{lukoff2020ancient} to scaffold the practice and help individuals build independence. The system could provide recommendations for alternative resources when a user cannot find a suitable live session, such as when the system detects that a user is frequently switching meditation sessions. 

\textit{Incorporate journaling}. 
We found that viewing others sharing their personal stories in live meditation gave us a sense of connection to them, and sharing our own thoughts helped us to become more engaged in the session and reflect on our experiences. 
Additionally, while we used journaling for the purpose of documenting our autoethnographic study, it provided us with an opportunity to reflect on our inner meditative process, benefits, and obstacles. Looking back, this journaling was useful to reflect on the role that meditation practice played in our mental well-being after the three-month study ended. 
In line with research showing that journaling is associated with improved mental well-being \cite{Smyth2018-wu}, we suggest that a live meditation system could include reflection journaling features (similar to those in \cite{Pirzadeh2013-eg}), such as sending out writing prompts to meditators after a session to help them track their mindfulness journey. 

\textit{Make it social}. 
As discussed above, social interactions in live meditation sessions motivated us and the teachers we interviewed to commit to our practice and build a more consistent meditation practice routine. 
Further, in our autoethnographic study, we found it useful to attend live small-group meditations organized by fellow meditators with recorded soundtracks, allowing us to practice and have informal chats with others. Such shared experiences are useful when the teacher is not available, or when one seeks companionship rather than active guidance \cite{Lee2021-study_with_me, Taber2020-companion}, or support from a small group for their mental or spiritual practice \cite{almedom2005social}. Given that research has shown a clear positive relationship between having social connections and mental health \cite{Kim2022-xy}, we believe that any mental well-being tool, technology, practice, or program would benefit from connecting individuals to others.

\subsection{Meditation Live Streams Vs. Other Mediums Vs. Other Genres of Live Streams}
\begin{table*}
\caption{A summary of findings for RQ3: The differences between live-stream meditation and other forms.}
\resizebox{\textwidth}{!}{%
\begin{tabular}{lll}
\toprule
\textbf{Medium} & \textbf{Genre}
& \textbf{How live-stream meditation is different} \\ \hline
Pre-recorded Soundtracks & Meditation
& More social interaction  \\
& & More personal stories \\
& & More motivation to practice  \\ 
\hline
In-person sessions & Meditation
& Lower cost (time, finance) \\ 
& & More options (variety of meditation) \\
& & More freedom to join \& leave \\
& & More anonymity \\
& & Larger scale \\ 
\hline
Live Streaming & Informal knowledge-sharing
& Mostly eye closed, listen and practice \\
& Fitness & Mostly sedentary, less physical activities \\ 
& Music Performance & Less influence from audience on stream content \\
& & Attention to self \\ 
& Companionship & Less pursuit in productivity \\
\bottomrule
\end{tabular}
\label{table_difference}
}
\end{table*}

As we discussed in Section \ref{bg:live-stream}, meditation live streams can be seen as a form of informal knowledge sharing and companionship. Our research found that live meditation shares characteristics with other live stream genres but also possesses its uniqueness. We summarize the differences in Table \ref{table_difference}.

Based on our findings, transitioning in-person meditation sessions to live streams offers tangible benefits, such as saving time on the commute and costs associated with physical venues, 
Live streams allow teachers to reach a more expansive audience, and provide participants with a wider array of sessions and the freedom to join or leave at their convenience.
Additionally, the interactive nature of live streams can foster more motivation than pre-recorded tracks. 
Despite these benefits, there are salient drawbacks. Live streams lack the constant availability of pre-recorded tracks and the physical presence and "collective energy" found in traditional in-person sessions \cite{Vieten2018-ma}. 

Based on our own experiences and observations in the chat, meditating alongside the streamer is a common practice in meditation live streams, which is similar to following instructions and doing movements in fitness streams \cite{Guo2022-fitness}. 
They both focus on applying the techniques in practice instead of gaining intellectual knowledge. 
One difference lies in the streamer and attendees' relationship and physical distance to the screen. 
In fitness streams, people move their bodies a few feet away from the screen, making it difficult to provide and receive real-time feedback \cite{Guo2022-fitness}. 
In contrast, most live stream meditative activities are sedentary, so responding through the streaming interface is more easily accessible. However, the meditation practice itself calls for tuning out screens and devices, resulting in tension between the streaming activity and the meditation activity. 

Some teachers broadcast meditative music instead of verbally guiding the audience to practice meditation. They utilize technologies such as a stream deck, multiple cameras, and lighting to augment the viewers' experience. 
These streams are similar to live music performances \cite{Thomas2020-live-performance-music}, but are quieter, more peaceful, and involve less streamer-viewer engagement, such as frequently requesting songs or reading and writing comments. 
As such, live meditation viewers usually do not actively participate or affect the streamed content. Most of the time they are guided to enjoy the music with their eyes closed and focus on their inner sensations.

Like other genres, donation is an important feature in meditation live streams. Teachers reported that the monetary incentives encouraged them to continue offering live sessions. 
But unlike other entertainment-oriented streams \cite{Wang2019-greedy, Li2019-co-performance, Lee2019-gifts}, the teachers we interviewed felt conflicted about requesting donations, and they did not report, nor did we observe, that the audience used tipping as a strategy to grab the teacher's attention or request personalized content.

We found music-based meditation or silent meditation without verbal guidance echoes companionship and "study with me" (SWM) streams \cite{Taber2020-companion, Lee2021-study_with_me} in terms of communication patterns and viewers' behaviors. 
In both cases, the role of the streamer is deemphasized, and audience members use the stream to create an external ambience for meditation, study, or work. 
However, viewers of SWM videos pursue concentration and extended study time \cite{Lee2021-study_with_me}, whereas our motivation for attending meditation live streams was to build a practice routine and to interact with the mindfulness community.

\subsection{Redesigning Live Stream Services for Meditation}
\label{discussion:redesign}

Given the unique challenges that we, as meditators, and the teachers faced in live meditation, a traditional live stream tool may not sufficiently support the practice sufficiently. Based on our findings, we propose a few recommendations for redesigning live stream services for meditation.

\begin{figure}[hbt]
    \includegraphics[width=\hsize]{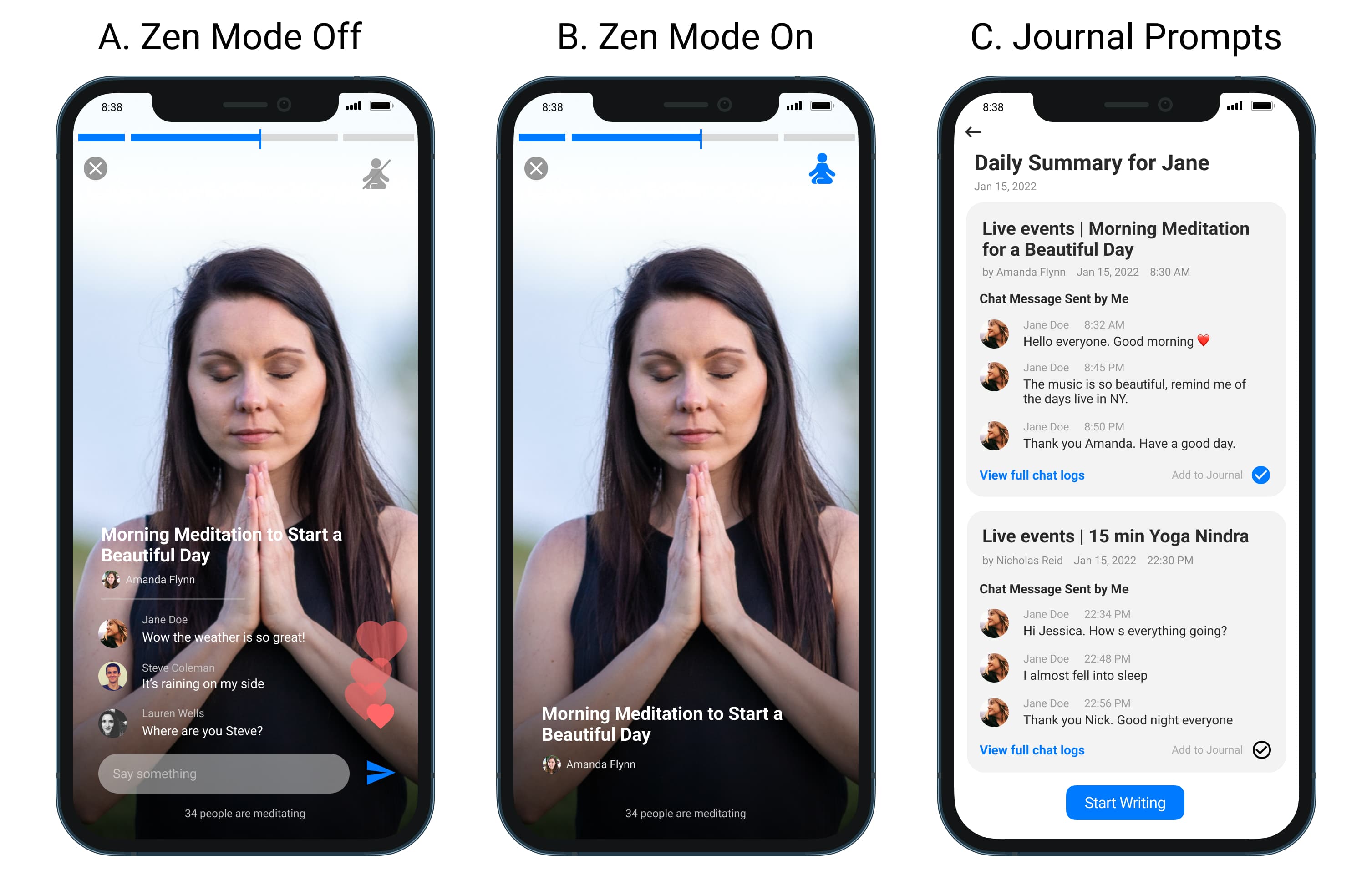}
    \caption{Design ideas for meditation live streams.
    A: Zen mode is off. The audience can send hearts or chats as they do in existing live stream platforms.
    B: Zen mode is on. Interaction features are disabled.
    C: Live events chat log as journal prompts. Users can review the live events they attended and the chat messages they sent before writing journal entries. 
    Notes: Meditator photo in A and B by Benjamin Child on Unsplash. User names are randomly generated.}
    \label{fig:design}
\end{figure}

\subsubsection{Non-distracting Live Meditation}
\label{design:non-distracting}

Our findings point to distractions from the live stream interface as a major challenge for both attendees and teachers. These findings are in line with previous findings that device notifications and constant connectivity are inversely correlated with inattention \cite{Kushlev2016Silence}.  
One way to address this is to add a "zen mode" where interactions are constrained during meditation practice. 
When turned on, attendees can only send messages to the teacher but are not able to chat with peer meditators, send hearts, or send virtual gifts. 
After the meditation practice, "zen mode" is turned off, and all interactions can resume.
Figure \ref{fig:design} A and B display example interfaces of "zen mode" on and off.
In addition to being set the "zen mode" on by the teacher, we think it would be useful as an option for attendees to set. 
This is because people join live meditation with different motivations, for example, to practice, to connect with others, or just to explore. Attendees who are not in "zen mode" would have access to all live stream features such as chat and hearts.

\subsubsection{Structured Live Sessions}
In our autoethnographic study, we found that live meditation services lack clear information about the session, such as techniques, agenda, and current status. From the teachers' end, we learned that they follow a standard script for creating live sessions, depending on the platform. These findings suggest that a more comprehensive system for providing information about sessions could be helpful in assisting meditators in finding suitable sessions. 
The system would allow teachers to input their agenda and expected timing, for example, two minutes for greetings and introduction, 15 minutes for meditation practice, and 10 minutes for Q\&A. This would allow viewers who join the session after it has started to orient themselves and know what to expect. 
Further, previous work has proposed manual or automatic segmentation of archived live streams \cite{Lu2018-streamwiki, Fraser2019-creative}. 
With teachers' input and audio recognition techniques, automatic live segmentation may become feasible in live stream meditation. 
Figure \ref{fig:design} A and B show an example interface with a progress bar at the top. The example live session has three sections and it is in the middle of the second section. Users can interact with the progress bar to see detailed information such as section descriptions.

\subsubsection{Journal Prompts with Live Chat Summary}

Incorporating journaling into live meditation is one practice we suggested for utilizing live streams as a resource for mental well-being. 
Journaling helped us reflect on our meditation journeys, requiring our intention and capacity to record our experiences. To support this process, we suggest that a live stream system could provide journal prompts with a summary of the live chat to inspire the audience to reflect on their practice after the session. 
A summary of the session would also serve as a qualitative and meaningful record of the meditation practice rather than providing numeric data (e.g., minutes, consecutive days of practice), which has been criticized in mobile mindfulness apps \cite{lukoff2020ancient}.
Figure \ref{fig:design}C shows an example interface of journal prompts from live chats. Users can review live events they attended and see chat messages they sent, which may trigger reflection on the thoughts and feelings they had during the live session. Users could import their chat messages into a journal and write reflections based on the logs. Given privacy considerations, we do not recommend sharing other attendees' chat messages as we observed that some meditators share very personal stories.  

\subsubsection{A More Well-being Oriented Live Stream Ecosystem}

Based on our findings, we think that live stream platforms should pay special attention to meditation live streams to highlight their mental well-being benefits. 
One way to address this is to separate meditation streams from the existing ranking system, giving them more visibility and avoiding the use of monetary incentives as a metric of success, especially in a general live stream platform (e.g., Twitch, Instagram) where the ranking is largely based on the density of interactions. 
Given the fact that meditation live streams usually contain a large part of the non-interactive time and have a much smaller viewership, applying the existing ranking system will consistently marginalize meditation streams. 
In addition, many of the meditation teachers we interviewed rely on viewer donations as their primary source of income. Given the monetary reward mechanism in popular live stream platforms, previous studies suggest that financial gain may gradually dominate streamers' motivations to contribute content \cite{Wang2019-greedy, Lee2019-gifts}, which is incongruent with the underlying motivations of teachers offering live meditation practice. 
Changing the reward system to free teachers from the pressure of soliciting donations could prevent potential negative effects on their motivations \cite{Wang2019-greedy}, offerings \cite{Lee2019-gifts}, and burnout \cite{Lorenz2021-nyt-creator-burnout}.  

\section{Limitation and Future Work}
Our autoethnographic and interview study provided us with a comprehensive understanding of live meditation from both the attendees' and teachers' perspectives, aiming toward designing live stream technologies that better support mindfulness. 
However, we acknowledge that autoethnography, although a valued method in HCI research \cite{Cunningham2005Autoethnography}, inherently captures our subjective perspective, and playing the role of a user did not necessarily offer us a full understanding of the range of live meditation experiences. 
Further research could mitigate this limitation by conducting interviews, diary studies, and other ethnographic techniques with diverse groups of live meditation attendees. 
In particular, these methods could shed light on the variety of live meditation experiences among people with different backgrounds, meditation experience levels, and technological platforms.

Extending beyond the domain of meditation, our research offers a generalizable framework that can be applied to other contexts where audiences engage with individual practice during live stream events, such as cooking, knitting, and studying. 
Moreover, it is also relevant in scenarios that involve self-reflective activities, including journaling and spiritual exercises like praying.
Furthermore, like previous work that investigated niche domains, our research aims to inspire both academic researchers and industry practitioners to undertake a critical examination of the unique characteristics and behaviors of users in various contexts. It also opens the discussion on whether standard design elements commonly found in mainstream genres are indeed appropriate or required in more specialized domains.

\section{Conclusion}
In this paper, we present findings from a three-month autoethnographic study and semi-structured interviews with 10 meditation teachers of live stream meditation. Our contribution is twofold, (1) an in-depth understanding of motivations, interactions, and challenges of live meditation from both the teachers' and the audience's perspectives, and (2) recommendations for utilizing and redesigning live stream tools to support meditation practice for mental well-being. 
 
\begin{acks}
This work is supported by Cornell CCSS Qualitative and Interpretive Research Institute. We thank all reviewers for their valuable feedback, as well as all the live stream meditation teachers who participated in the interviews and shared their experiences. Additionally, we extend our thanks to Nayeon Kwon, Huong Pham, and Ryun Shim for their research assistance.
\end{acks}

\bibliographystyle{ACM-Reference-Format}
\bibliography{sample-manuscript}



\end{document}